\documentclass{sf2a-conf2015}
\usepackage{graphicx}
\usepackage{hyperref}
\usepackage[]{natbib}  
\usepackage{epstopdf}
\usepackage{floatrow}

\def\BibTeX{{\rm B\kern-.05em{\sc i\kern-.025em b}\kern-.08em
    T\kern-.1667em\lower.7ex\hbox{E}\kern-.125emX}}
\bibpunct{(}{)}{;}{a}{}{,}  

\begin{document}

\TitreGlobal{SF2A 2015}


\title{Shocks, star formation and the JWST}

\runningtitle{Shocks, star formation and the JWS}

\author{A. Gusdorf}\address{LERMA, Observatoire de Paris, PSL Research University, CNRS, UMR 8112, F-75014, Paris, France; Sorbonne Universit\'es, UPMC Univ. Paris 6, UMR 8112, LERMA, F-75005, Paris, France}

\setcounter{page}{237}


\maketitle


\begin{abstract} The interstellar medium (ISM) is constantly evolving due to unremitting injection of energy in various forms. Energetic radiation transfers energy to the ISM: from the UV photons, emitted by the massive stars, to X- and $\gamma$-ray ones. Cosmic rays are another source of energy. Finally, mechanical energy is injected through shocks or turbulence. Shocks are ubiquitous in the interstellar medium of galaxies. They are associated to star formation (through jets and bipolar outflows), life (via stellar winds), and death (in AGB stellar winds or supernovae explosion). The dynamical processes leading to the formation of molecular clouds also generate shocks where flows of interstellar matter collide. Because of their ubiquity, the study of interstellar shocks is also a useful probe to the other mechanisms of energy injection in the ISM. This study must be conducted in order to understand the evolution of the interstellar medium as a whole, and to address various questions: what is the peculiar chemistry associated to shocks, and what is their contribution to the cycle of matter in galaxies ? What is the energetic impact of shocks on their surroundings on various scales, and hence what is the feedback of stars on the galaxies ? What are the scenarios of star formation, whether this star formation leads to the propagation of shocks, or whether it is triggered by shock propagation ? What is the role of shocks in the acceleration of cosmic rays ? Can they shed light on their composition and diffusion processes ? In order to progress on these questions, it is paramount to interpret the most precise observations with the most precise models of shocks. From the observational point of view, the James Webb Space Telescope represents a powerful tool to better address the above questions, as it will allow to observe numerous shock tracers in the infrared range at an unprecedented spatial and spectral resolution.


\end{abstract}

\begin{keywords}
   shock waves -- 
   astrochemistry --
   stars: formation --
   ISM: jets and outflows --
   ISM: kinematics and dynamics --
   Infrared: ISM
\end{keywords}

\begin{figure}
\floatbox[{\capbeside\thisfloatsetup{capbesideposition={left,top},capbesidewidth=0.4\textwidth}}]{figure}[\FBwidth]
{\caption{The ubiquity of shocks in the interstellar medium. Note the sizes indicated in the various panels. {\bf a):} The BHR71 bipolar outflow system driven by two young proto-stars, seen in colours and red contours at 8~$\mu$m by \textit{Spitzer}/IRAC, and in white contours in the CO (3--2) emission line by the APEX telescope; adapted from \citet{Gusdorf151}. {\bf b):} The W43-MM1 ridge, seen in colours and grey contours in the SiO (2--1) transition (adapted from \citealt{Nguyenluong13}). The white contours are column density contours. The white hexagons mark the position of massive dense cores identified by \citet{Louvet14}. The red hexagon is a dense core. {\bf c):} The IC 443 supernova remnant, seen in the CO (6--5) (colours) and (2--1) (white contours) transitions. The typical beam size of the CO observations is also shown, as well as those of current (HESS) and future (CTA) $\gamma$-ray observatories (Gusdorf et al., prep., I).}\label{figure1}}
{\includegraphics[width=0.6\textwidth]{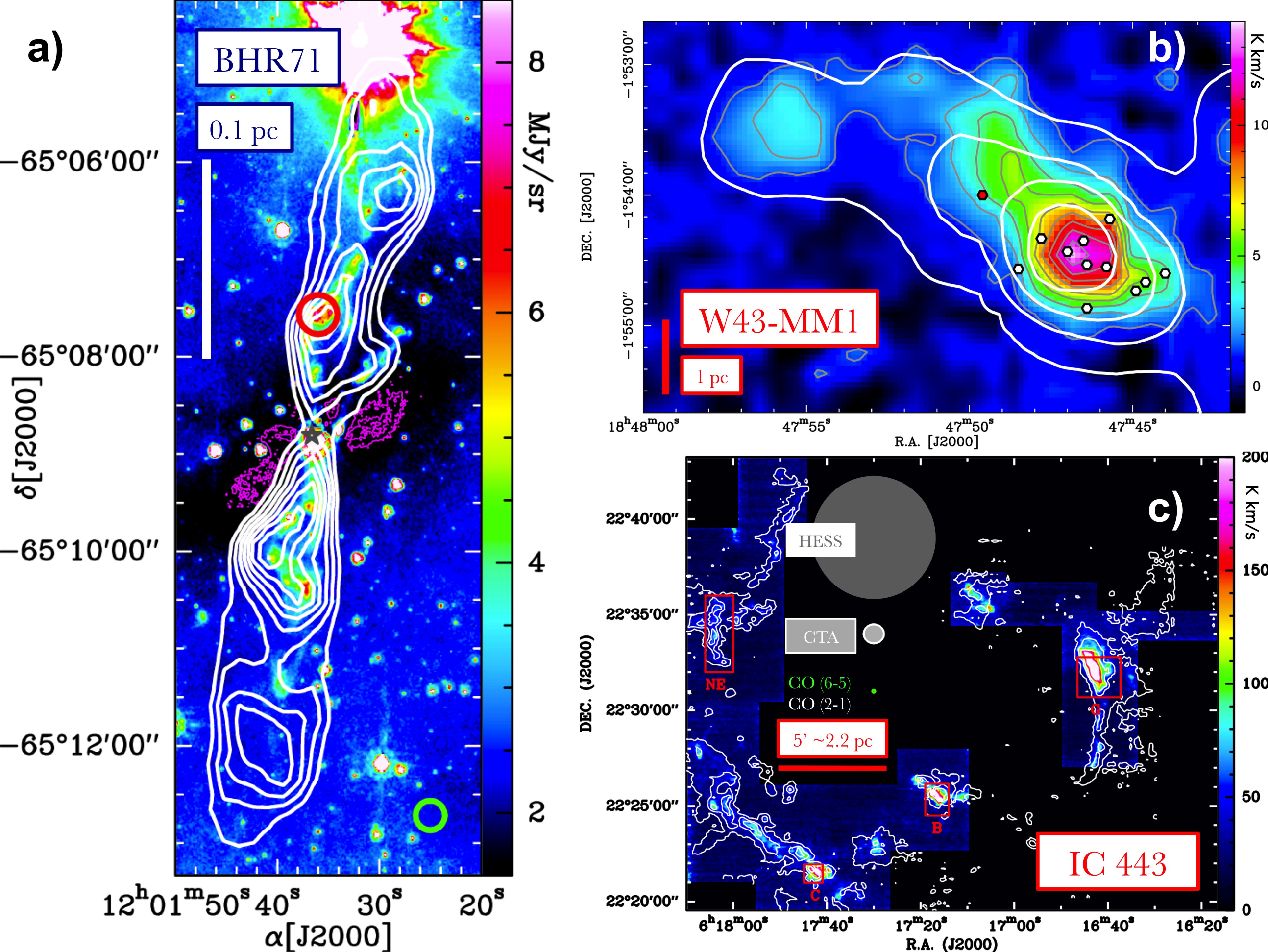}}
\end{figure}

\section{Different shock environments}

Observations over the past few decades have shown that, in the early stages of low-mass star formation, the process of mass accretion is almost always associated with mass ejection in the form of collimated jets. The jets impact on the parent cloud, driving a shock front through the collapsing interstellar gas. Large cavities, called bipolar outflows, are carved in the ambient medium, which is accelerated, compressed and heated by the shock wave. This paradigm was proposed by \citet{Snell80} and has been regularly verified in such environments (see, for example, figure~\ref{figure1}, and \citealt{Arce07, Frank14} for reviews), including recent high-angular-resolution observations by ALMA (e.g. \citealt{Codella14}). Comparisons of observations in these jet/outflow regions with 1D shock models such as the Paris-Durham model \citep{Flower15} have been extensively performed in the last decade. The aim is often to constrain the physical conditions (e.g. \citealt{Giannini04, Giannini06, Dionatos10}). It can additionnally be to understand the chemistry of a given species, such as SiO (e.g. \citealt{Gusdorf081,Gusdorf082, Gusdorf151}) or/and CH$_3$OH and NH$_3$ (e.g. \citealt{Flower102, Flower12}). Of course, when properly understood, the chemistry of this species can in turn allow to better characterize the physical conditions prevailing in the shocked gas. Recently, the column densities of H$_2$O inferred from observations in various low-mass star forming regions have raised some concern that models (e.g. \citealt{Kaufman96,Flower101,Gusdorf11}) might have overestimated its abundance. This discrepancy could be due to the underestimate of the effects of the protostellar radiation field in models, but this assumption has not been tested yet. In any case, directly comparing observed line integrated intensities to modelled ones does not seem to yield such a discrepancy (e.g. \citealt{Flower13,Neufeld14}) in low-mass star forming environments. 

One of the major shortcomings of such models is their simple geometry. In the case of the Paris-Durham model, the choice was made to keep a 1D geometry in order to optimize the numerical treatment of microscopic physical and chemical processes, as a 3D shock model including 1000 chemical reactions would still take too much time to run. Recently, attempts have been made to overcome this difficulty. In \citet{Kristensen08} and \citet{Gustafsson10}, the 1D shock models have been stitched on 2D or 3D structures, ultimately and successfully allowing to compare modelled maps of integrated intensity with observations. The Paris-Durham model has also been combined with dynamical MHD disk wind models in order to account for the observations of molecular counterparts to atomic jets, very close to the protostars and at very early evolutionary stages (\citealt{Panoglou12} and Yvart et al., in prep.). Probability density functions of 1D shock layers have also been generated in order to statistically account for a more complex shock structure (\citealt{Lesaffre13}).

Interestingly, these latest shock models somehow included the effects of the protostellar radiation field. Such an inclusion is paramount when studying shocks in regions of more massive star formation, as massive protostars radiate significant amounts of energy in the UV range (see \citealt{Tan14} for a review). Mostly because of this extra ingredient, the paradigm for the formation of massive stars is less established than for their low-mass counterpart. The key question is to determine whether massive stars form from a dense core in a scaled-up version of the low-mass scenario, or if alternative processes should be invoked, such as competitive accretion or coalescence of low-mass protostars. A way to build a consistent view on star formation is to compare observations with models in regions of shocks associated to the formation of protostars of various masses. This implies a development of a new class of shock models, called \lq irradiated' because they must include the influence of the protostellar radiation field. Indeed, \citet{Lefloch15} have been able to successfully reproduce CO observations of a shock in an intermediate-mass protostellar outflow, but e.g. \citet{Leurini14}, \citet{Gusdorf152}, and Gusdorf et al., submitted to A\&A, have quantified the shortcomings of non-irradiated shock models to fit H$_2$O, OH, or C$^+$ observations in regions of massive star formation. The implementation of irradiated shock models is ongoing (e.g. \citealt{Lesaffre13}, \citealt{Melnick15}) and is key to constrain the scenarios of massive star formation, but also its chemical and energetic impact. It is also important to be able to distinguish regions where shocks are irradiated by an external radiation source from regions where the shocks itself is a source of UV radiation, a situation that occurs in fast shocks (e.g. \citealt{Hollenbach89}).

The implementation of the effects of a mild irradiation is also necessary to better understand the peculiar physics and chemistry at work in the dense filaments whose formation is the result of dynamical accretion processes themselves at the origin of the formation of molecular clouds. In these converging flows, dense and low-velocity shocks are observed (see figure~\ref{figure1} and e.g. \citealt{Nguyenluong13}, \citealt{Duartecabral14}). Studying these regions where streams of matter collide or converge is a way to understand the ongoing massive star formation, often clustered in some regions of the ridge \citep{Louvet14}. It is also a way to confront our models with a new chemistry where thermal effects can dominate the sputtering of grains, resulting e.g. in the detection of extended, and narrow emission of SiO transitions for instance (Louvet et al., in prep.). It is finally a way to quantify whether these extended and very common regions contribute to the energetic balance of galaxies, through measurements of the flux from their high-lying CO transitions (Gusdorf et al., in prep., II). 

The objective of quantifying the high-$J$ CO emission from shock regions, or more generally the energetic impact they could have on large scales, is also significant when studying supernova remnants (SNRs; see figure~\ref{figure1}), given their size. In such regions, various studies have shown that non irradiated models of shocks, stationary (\citealt{Gusdorf12}, \citealt{Neufeld14}) or not (\citealt{Cesarsky99}, \citealt{Anderl14}), can reproduce the observations of CO, H$_2$ and/or H$_2$O. In old SNRs, the detailed characterisation of the interstellar content can also serve to support the study of cosmic-ray (CR) related questions. Indeed, these objects are usually detected at very high energy by $\gamma$-ray telescopes such as HESS, \textit{Fermi}, VERITAS or MAGIC (e.g. \citealt{Aharonian08}, \citealt{Abdo10}). CRs have been accelerated to high energies in the past, then trapped in the shocked/dense region. The $\gamma$-ray emission is the signature of interactions involving the hadronic ($\pi^0$ decay) or leptonic (Inverse Compton, Bremstrahlung, synchrotron emission) part of CRs on the one hand, and the dense and often shocked medium in the other hand. Characterizing shocks is hence essential to understand the contribution of these processes to the high energy spectra, and to support the study of CRs acceleration, composition and diffusion. 

\section{The JWST contribution}

With an expected launch in 2018, the JWST will be a powerful tool to support the study of shocks. First, it will allow to observe pure rotational line emission from H$_2$. H$_2$ is an important molecule in the ISM and particularly in shocked regions. Indeed, it is an abundant molecule, a key partner in chemical and collisional processes leading to the formation and excitation of other species. Since its rotational levels lie at a few hundreds to a few thousands of K, its excitation traces the dense, warm medium, and H$_2$ hence constitutes an important cooling agent of shocked regions. Constraining the physical conditions prevailing in the H$_2$ emitting gas is an unavoidable step when performing detailed chemical studies, and represent a potential benefit to every study already mentioned. This is all the more true in regions that were not observed by the ISO or \textit{Spitzer} telescopes. This is the case of the filaments and ridges that constitutes one of the important legacies of the \textit{Herschel} telescope and whose studies started when no further observation of rotational line emission from H$_2$ was possible with \textit{Spitzer}. Observing the H$_2$ emission in various environments with MIRI will also allow to better understand the processes of its formation at the surface of grains, and to place tighter constraints on its excitation processes. From the point of view of H$_2$ and more generally, the gain in angular resolution of the JWST with respect to those afforded by ISO, or the \textit{Spitzer} telescopes should allow to resolve the shock structures. Simultaneously the gain in sensitivity will allow to map larger regions on bipolar outflows our supernova remnants, an essential step when assessing the energetic impact of an entire object on its surroundings. Such detailed observations will also provide strong constraints for multi-dimensional shock modelling. This gain in sensitivity should also allow to study the propagation of shocks in the more diffuse medium, where H$_2$ emission is expected to be fainter. 

Insight will also be gained from the JWST in terms of chemistry, as a lot of filters have been designed to target specific lines or features (H$_2$, CO, H$_2$O ice, CH$_4$, CO, CO$_2$,...). When it comes to shocks, the detailed characterization of the ionization state of the gas is key to investigate the nature of shocks. In particular, the observation of atomic and ionized species, such as [Ar II], [Ne II], [Ne III], [S III], [S~I], [Fe II], [Fe III], [O I] should allow to quantify the dissociative nature of shocks, and the degree or origin of their irradiation. The observations of H$_2$O and the product of its dissociation such as OH and [OI] should also contribute to understand their potential to dissociate the ISM (see, e.g. \citealt{Tappe08,Tappe12} on the use of \textit{Spitzer} observations of OH to probe the potential of 40 km~s$^{-1}$ shocks to generate self-irradiation). Incidently, the ionization fraction reached in a shock is also a key parameter when it comes to assessing its potential to accelerate cosmic rays (Padovani et al., subm. to A\&A). Moreover, the observation of numerous molecules (HD, H$_3$O$^+$, CH$_4$, HCN,...) by the JWST will allow to address more specific chemical questions. Finally, the JWST will allow to probe the composition of ice mantles, yielding new constraints on the formation paths of complex molecules. 


\section{Perspectives}
Beyond its unquestionable intrinsic value, the JWST will arrive at a unique time in astronomy, where windows of unprecedented quality are open at all the wavelength of the electromagnetic spectrum. Its strength will not only reside in its added value, but also in its complementarity with other observatories. Used with the VLT (and high resolution receivers such as CRIRES or VISIR), the JWST will provide the most comprehensive view on H$_2$ emission in terms of spectral and spatial resolutions and mapping capacities. Beyond infrared diagnoses, there will be a natural complementarity of the JWST with sub-mm to far infrared telescopes such as the IRAM-30m, APEX, \textit{Herschel} or SOFIA, to investigate the chemistry of the ISM in greater details than ever. This complementarity extends towards the high energies, as the combination of the JWST with X- (ATHENA) and $\gamma$-ray (HESS, \textit{Fermi}, MAGIC, VERITAS, and more importantly CTA) telescopes will allow to better describe fast shocks and to improve our understanding of CRs related questions. Finally, shock astrophysics will benefit from the combined use of the JWST with ALMA, that will allow to probe the effect of shocks not only in our own Galaxy, but in all galaxies, thus tracing the shocks effects in the history of the Universe (see P. Guillard's contribution).

\bibliographystyle{aa}  
\bibliography{sf2a-template} 

\end{document}